\documentclass[conference]{IEEEtran}

\usepackage{graphicx}
\usepackage{caption}
\usepackage{subcaption}
\usepackage{color}
\ifCLASSINFOpdf
\else
\fi

\IEEEoverridecommandlockouts
\usepackage{mathtools}
\usepackage{flushend}
\usepackage{amsmath}
\usepackage{algorithmic,algorithm}
\usepackage{cite}

\allowdisplaybreaks

\begin{document}

\title{Throughput Analysis for Relay-Assisted Millimeter-Wave Wireless Networks}

\author{\IEEEauthorblockN{Cristian Tatino\IEEEauthorrefmark{1}\IEEEauthorrefmark{2}, Nikolaos Pappas\IEEEauthorrefmark{1}, Ilaria Malanchini\IEEEauthorrefmark{2}, Lutz Ewe\IEEEauthorrefmark{2}, Di Yuan\IEEEauthorrefmark{1}
\thanks{This project has received funding from the European Union's Horizon 2020 research and innovation programme under the Marie Sklodowska-Curie grant agreement No. 643002.}}
\IEEEauthorblockA{\IEEEauthorrefmark{1}Department of Science and Technology, Link\"{o}ping University, Sweden\\
               Email: \{cristian.tatino, nikolaos.pappas, di.yuan\}@liu.se}      
               \IEEEauthorblockA{\IEEEauthorrefmark{2}Nokia Bell Labs, Stuttgart, Germany\\
               Email: \{ilaria.malanchini, lutz.ewe\}@nokia-bell-labs.com}
}

\maketitle
\IEEEpeerreviewmaketitle
\begin{abstract}
In this work, we analyze the throughput of random access multi-user relay-assisted millimeter-wave wireless networks, in which both the destination and the relay have multi-packet reception capability. We consider a full-duplex network cooperative relay that stores the successfully received packets in a queue, for which we analyze the performance. Moreover, we study the effects on the network throughput of two different schemes, by which the source nodes transmit either a packet to both the destination and the relay in the same timeslot by using wider beams (broadcast scheme) or to only one of these two by using narrower beams (fully directional scheme). Numerical results show how the network throughput varies according to specific system parameters, such as positions and number of nodes. The analysis allows us also to understand the optimal transmission scheme for different network scenarios and shows that the choice to use transmissions with narrow beams does not always represent the best strategy, as wider beams provide a lower beamforming gain, but they allow to transmit simultaneously both at the relay and the destination.
\end{abstract}

\section{Introduction}
\label{sec:Intro}

Given the exponential growth of data rate and connections for the fifth generation (5G) of wireless networks, millimeter-wave (mm-wave) communications technology has attracted the interest of many researchers in the past few years. The abundance of spectrum resource in the mm-wave frequency range (30-300 GHz) could help to deal with the longstanding problem of spectrum scarcity.
However, the signal propagation in the mm-wave frequency range is subject to more challenging conditions in comparison to lower frequency communications, especially in terms of path loss and penetration loss, which causes frequent communication interruptions.

Several solutions have been proposed in order to overcome the blockage issue, e.g., cell densification, multi-connectivity and relaying techniques. Although relay has been extensively investigated for microwave frequencies~\cite{Coop1,Coop2,Rel1,Niko,Rel2}, mm-wave communications present peculiarities that make further analysis necessary. As an example, in contrast to broadcast transmissions (mainly used for lower frequency bands), mm-waves use narrow beams with higher beamforming gain to overcome the path loss issue. By using these transmissions (fully directional scheme, $\mathrm{FD}$), a source node (user equipment, UE) sends a packet either to the relay or to the destination  (mm-wave access point, mmAP). On the other hand, in the broadcast communication case ($\mathrm{BR}$), a packet that is sent by a UE can be received by both the relay and the mmAP in the same timeslot. 

In this work, we analyze the throughput of network cooperative communications in a multi-user mm-wave wireless network. We evaluate two types of transmissions, i.e., $\mathrm{FD}$ and $\mathrm{BR}$. When the UEs use a $\mathrm{BR}$ scheme and the transmission to the destination fails, the relay stores the packets (that are correctly decoded) in its queue and is responsible to transmit it to the destination. This technique is also known as network level cooperation relaying~\cite{Coop2, Rel1, Niko, Rel2}.
\subsection{Related Work}
\label{sec:Rel}
The benefits of relaying techniques for mm-wave wireless networks have been discussed in several works, e.g,~\cite{RelayPhy1,RelayPhy2,RelayBlo,FairRelay,ProbD2D,CovD2D,FallRelay}. In~\cite{RelayPhy1} and \cite{RelayPhy2}, stochastic geometry is used to analyze the system performance for a relay-assisted mm-wave cellular network. Authors analyze several relay selection techniques and they show a significant improvement in terms of signal-to-interference-plus-noise ratio (SINR) distribution and coverage probability. In~\cite{RelayBlo}, authors propose a two-hop relay selection algorithm for mm-wave communications that takes into account the dependency between the source-destination and relay-destination paths in terms of line-of-sight (LOS) probability. In~\cite{FairRelay}, a joint relay selection and mmAP association problem is considered. In particular, the authors propose a distributed solution that takes into account the load balancing and fairness aspects among multiple mmAPs.
Other works, \cite{ProbD2D} and~\cite{CovD2D}, focus on relaying techniques for device-to-device (D2D) scenarios and analyze, by using stochastic geometry, the coverage probability and the relay selection problem, respectively.

The authors of~\cite{FallRelay} analyze the tradeoff between mm-wave relay and microwave frequency transmissions for a two-hop half-duplex relay scenario. They study the throughput and delay for a single source-destination pair and a single relay, which can transmit on mm-wave frequencies or by using microwave frequencies when the direct path is blocked.
To the best of our knowledge, the setup considered in this paper has been investigated only for microwave frequencies ~\cite{Niko}, without taking into account different transmission strategies.

\subsection{Contributions}
\label{sec:Contributions}
In this work, we provide an analysis of the throughput for random access multi-user cooperative relaying mm-wave wireless networks. We consider two different transmission schemes, i.e., $\mathrm{FD}$ and $\mathrm{BR}$ that may provide different beamforming gains and cause different interference levels. Indeed, $\mathrm{BR}$ transmissions may provide a lower beamforming gain with respect to the $\mathrm{FD}$ scheme, but they allow to transmit simultaneously both at the relay and the mmAP. The UEs, independently, choose to transmit by following one of the schemes and we identify the optimal strategy with respect to system parameters; namely, we show under which conditions $\mathrm{BR}$ transmissions should be preferred to a $\mathrm{FD}$ scheme and vice-versa. Furthermore, by using queueing theory, we study the performance characteristic of the queue at the relay, for which we derive the stability condition, as well as the service and the arrival rate. 

The rest of the paper is organized as follows: in Section \ref{sec:Ass}, we describe the system model and the assumptions. In Section~\ref{sec:PA}, we present the queue analysis at the relay with two UEs and in Section~\ref{sec:Thr}, we generalize these results and evaluate the aggregate network throughput for $N$ UEs. 
In Section~\ref{sec:Res}, we illustrate the results and performance evaluation and Section~\ref{sec:Conc} concludes the paper.

%
%
%

\section{System Model and Assumptions}
\label{sec:Ass}

\subsection{Network Model}
\label{sec:NM}
We consider a set of symmetric\footnote{Symmetric UEs have the same mm-wave networking characteristics, e.g., propagation conditions. Our study can be generalized to the asymmetric case; however, the analysis will be dramatically involved without providing any additional meaningful insights.} UEs $\mathcal{N}$, with cardinality $N$. We consider one mmAP (destination) and one full-duplex relay ($R$) that operates in a decode-forward manner. We assume multiple packet reception capability both at the mmAP and the $R$ which are equipped with hybrid beamformers and they can form multiple beams at the same time \cite{Hybrid}. The UEs are equipped with analog beamformers, which can form one beam at a time. We assume slotted time and each packet transmission takes one timeslot. The relay has no packets of its own, but it stores the successfully received packets from the UEs in a queue, which has infinite size\footnote{A similar analysis can be derived for the case of finite queue size, which will be treated in an extension of this work.\label{infinte}} and bursty arriving traffic. The UEs have saturated queues, i.e, they are never empty. We assume that acknowledgments (ACKs) are instantaneous and error free and packets received successfully are deleted from the queues of the transmitting nodes, i.e., UEs and $R$. 

\begin{figure}[tb]
	\centering
	\includegraphics[width=5.5cm]{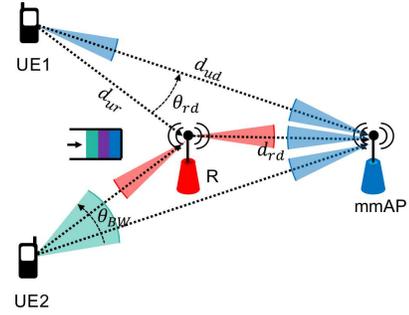}
	\caption[]{$\mathrm{FD}$ (UE1) and $\mathrm{BR}$ (UE2) transmissions in a scenario with two UEs, one relay and one mmAP. In this example, UE1 is transmitting to the mmAP.}
	 \label{fig:FB}
\end{figure}

UEs and $R$ transmit a packet with probabilities $q_{u}$ and $q_{r}$, respectively. As mentioned previously, a UE can transmit by using either the $\mathrm{FD}$ or the $\mathrm{BR}$ scheme with probabilities $q_{uf}$ and $q_{ub}$ ($q_{uf}=1-q_{ub}$), which are conditioned to the event that a packet is transmitted. In turn, when a UE uses the $\mathrm{FD}$ transmission, it transmits either to the mmAP or to $R$ with probabilities $q_{ud}$ and $q_{ur}$ ($q_{ud}=1-q_{ur}$), respectively, which are conditioned probabilities to the event that a packet is transmitted by using the $\mathrm{FD}$ scheme. In the $\mathrm{BR}$ case, $R$ stores the successfully received packets only when these are not received by the mmAP and the relay always uses directional communications to forward them to the mmAP. 
In Fig.~\ref{fig:FB}, we illustrate an example of the $\mathrm{FD}$ and $\mathrm{BR}$ transmissions, where $d_{ur}$ and $d_{ud}$ represent the distances between the UE and $R$ and between the UE and the mmAP, respectively. The parameter $\theta_{rd}$ is the angle formed by $R$ and the mmAP with a UE as vertex and $\theta_{BW}$ is the beamwidth. Hereafter, we indicate the probability of the complementary event by a bar over the term (e.g., $\overline{q}_{u}=1-q_{u}$). Moreover, we use superscripts $f$ and $b$ to indicate the $\mathrm{FD}$ and $\mathrm{BR}$ transmissions, respectively. 

\subsection{SINR Expression and Success Probability}
\label{sec:SINRexp}
A packet is successfully received if the SINR is above a certain threshold $\gamma$. Ideally, multiple transmissions at the receiver side of a node do not interfere when they are received on different beams. However, in real scenarios, interference cancellation techniques are not perfect; thus, we introduce a coefficient $0\le \alpha \le1$ that models the interference between received beams. The cases $\alpha=0$ and $\alpha=1$ represent perfect interference cancellation and no interference cancellation, respectively. In order to keep the clarity of the presentation we consider $\alpha$ constant. Moreover, we assume that an $\mathrm{FD}$ transmission to the mmAP does not interfere with the packet transmitted to $R$ and vice-versa. On the other hand, when a UE uses a $\mathrm{BR}$ scheme, its transmission interferes with the transmissions of the other UEs for both the mmAP and $R$. 
\begin{figure*}[!t]
\normalsize
\begin{equation}
\label{eq:SINR}
\begin{aligned}    
&\text{SINR}_{ij/\mathcal{I}_{fl},\mathcal{I}_{fn},\mathcal{I}_{bl},\mathcal{I}_{bn}}^{f} |\text{LOS}_{ij}=\frac{p_{t}g_{i}^{f}g_{j}^{f}h_{l}(i,j)}{p_{N}+\alpha \Bigl(\displaystyle \sum_{k \in \mathcal{I}_{fl}}p_{r/l}^{f}(k,j)+\sum_{m \in \mathcal{I}_{bl}}p_{r/l}^{b}(m,j)+\sum_{u \in \mathcal{I}_{fn}}p_{r/n}^{f}(u,j)+\sum_{v \in \mathcal{I}_{bn}}p_{r/n}^{b}(v,j)\Bigl)}.
\end{aligned} 
\end{equation}
\hrulefill
\end{figure*}

We assume that the links between all pairs of nodes are independent and can be in two different states, LOS and non-line-of-sight (NLOS). Specifically, $\text{LOS}_{ij}$ and $\text{NLOS}_{ij}$ are the events that node $i$ is in LOS and NLOS with node $j$, with associated probabilities $P(\text{LOS}_{ij})$ and $P(\text{NLOS}_{ij})$, respectively. 
Furthermore, we assume that $R$ is placed in a position that guarantees the LOS with the mmAP, namely, $P(\text{LOS}_{rd})=1$. In order to compute the SINR for link $ij$, we first identify the sets of interferers that use $\mathrm{FD}$ and $\mathrm{BR}$ transmissions, which are $\mathcal{I}_{f}$ and $\mathcal{I}_{b}$, respectively. Then, we partition each of them into the sets of nodes that are in LOS and NLOS with node $j$. These sets are $\mathcal{I}_{fl}$ and $\mathcal{I}_{fn}$, for the nodes that use the FD scheme and $\mathcal{I}_{bl}$ and $\mathcal{I}_{bn}$ for the UEs that use the BR transmissions. 
Therefore, when node $i$ is in LOS with node $j$, we can write the SINR, conditioned to $\mathcal{I}_{fl},\mathcal{I}_{fn},\mathcal{I}_{bl},\mathcal{I}_{bn}$, as in~\eqref{eq:SINR}.

The beamforming gain of the transmitter and the receiver are $g_{i}$ and $g_{j}$, respectively. These are computed in according to the ideal sectored antenna model \cite{Bai}, which is given by: $g_{i}=g_{j}=\frac{2\pi}{\theta_{BW}}$ in the main lobe, and $0$ otherwise.
The term $h_{l}(i,j)$ is the path loss on link $ij$ when this is in LOS. The transmit and the noise power are $p_{t}$ and $p_{N}$, respectively. The terms $p_{r/l}(i,j)$ and $p_{r/n}(i,j)$ represent the received power by node $j$ from node $i$, when the first is in LOS and NLOS, respectively. Note that similar expressions of the SINR can be derived also in case of $\mathrm{BR}$ and NLOS. Finally, the success probabilities for a packet sent on link $ij$ by using $\mathrm{FD}$ and $\mathrm{BR}$ transmissions are represented by the terms $P_{ij/\mathcal{I}_{f},\mathcal{I}_{b}}^{f}$ and $P_{ij/\mathcal{I}_{f},\mathcal{I}_{b}}^{b}$, respectively. Here, we consider only the conditioning on the sets $\mathcal{I}_{f}$ and $\mathcal{I}_{b}$, since we average over all possible scenarios for the LOS and NLOS link conditions. The expression for the $\mathrm{FD}$ transmission and $N$ UEs is given in Appendix A. 

\section{Performance Analysis for the Relay Queue}
\label{sec:PA}
In order to compute the network throughput, in this section, we evaluate the arrival rate, $\lambda$, for the queue at $R$, for which we further analyze the service rate, $\mu_{r}$, and the stability condition. Namely, we present the results for two UEs to give insights to understand the throughput analysis, which is generalized for $N$ UEs in Section \ref{sec:Thr}. First, similar to~\cite{Niko}, we compute $\lambda$ as follows:
\begin{align*} 
\lambda&=P(Q=0)\lambda_{0}+P(Q\neq0)\lambda_{1}\stepcounter{equation}\tag{\theequation}\label{eq:lambda},
\end{align*} 
where $\lambda_{0}$ and $\lambda_{1}$ are the arrival rates at $R$ when the queue is empty or not, which occur with probabilities $P(Q=0)$ and $P(Q\neq0)$, respectively. Namely, when the queue is not empty, $R$ may transmit and interfere with the other transmissions to the mmAP.
Therefore, by considering all the possible combinations for the two UEs scenario, where $R$ can receive at maximum two packets per timeslot, we can compute $\lambda_{0}$ and $\lambda_{1}$. 
Note that the definition of the sets $\mathcal{I}_{f}$ and $\mathcal{I}_{b}$ can be simplified since the UEs are symmetric. Therefore, it is sufficient to indicate the number of UEs that are interfering and whether $R$ is transmitting, i.e., we indicate with $\{|\mathcal{I}_{f}|,r\}^{f}$ and $\{|\mathcal{I}_{f}|\}^{f}$ the sets of interferers that use $\mathrm{FD}$ transmissions when $R$ is transmitting or not, respectively, and with $\{r\}^{f}$ the set of interferers when only the relay is transmitting. Therefore, we obtain:
\begin{align*}
\lambda_{0}&=2q_{u}\overline{q}_{u}q_{uf}q_{ur}P_{ur}^{f}+2q_{u}\overline{q}_{u}q_{ub}P_{ur}^{b}\overline{P}_{ud}^{b}\\
&+q_{u}^2q_{uf}^2q_{ur}^2q_{ur}^2\Bigl[2P_{ur/\{1\}^{f}}^{f}\overline{P}_{ur/\{1\}^{f}}^{f}\\
&+2\Bigl(P_{ur/\{1\}^{f}}^{f}\Bigl)^{2}\Bigl]+2q_{u}^{2}q_{uf}^{2}q_{ur}q_{ud}P_{ur}^{f}\\
&+2q_{u}^{2}q_{1f}q_{ub}q_{ur}\Bigl[P_{ur/\{1\}^{b}}^{f}\Bigl(1-P_{ur/\{1\}^{f}}^{b}\overline{P}_{ud}^{b}\Bigl)\\
&+\overline{P}_{ur/\{1\}^{b}}^{f}P_{ur/\{1\}^{f}}^{b}\overline{P}_{ud}^{b}+2\Bigl(P_{ur/\{1\}^{f}}^{b}\overline{P}_{ud}^{b}\Bigl)^{2}\Bigl]\\
&+2q_{u}^{2}q_{ub}q_{uf}q_{ud}P_{ur}^{b}\overline{P}_{ud/\{1\}^{f}}^{b}\\
&+q_{u}^{2}q_{ub}^{2}\Bigl[2P_{ur/\{1\}^{b}}^{b}\overline{P}_{ud/\{2\}^{b}}^{b}\Bigl(1-P_{ur/\{1\}^{b}}^{b}\overline{P}_{ud/\{1\}^{b}}^{b}\Bigl)\\
&+2\Bigl(P_{ur/\{1\}^{b}}^{b}\overline{P}_{ud/\{1\}^{b}}^{b}\Bigl)^{2}\Bigl],\stepcounter{equation}\tag{\theequation}\label{eq:lambda02}
\end{align*}
where, ${q}_{u}$, $q_{ub}$, $q_{uf}$, $q_{ud}$, and $q_{ur}$ are introduced in Section~\ref{sec:NM}. Similarly, we obtain that $\lambda_{1}=\overline{q}_{r}\lambda_{0}+q_{r}A_{r}$, whereas the service rate is $\mu_{r}=q_{r}B_{r}$. The terms $A_{r}$ and $B_{r}$ are given in Appendix B. Now, we derive the condition for the queue stability, which is used to determine the throughput. By applying the Loyne's criterion~\cite{loynes}, we can obtain the range of values of $q_{r}$ for which the queue is stable by solving the equation $\lambda_{1}=\mu_{r}$. Thus, we have that the queue at $R$ is stable if and only if $q_{r_{min}}<q_{r}\le1$, where $q_{r_{min}}$ is given by:
\begin{equation}\label{eq:qmin}
\begin{aligned} 
q_{r_{min}}=\frac{\lambda_{0}}{\lambda_{0}+B_{r}-A_{r}}.
\end{aligned} 
\end{equation}
\begin{figure}[tb]
	\centering
	\includegraphics[width=6cm]{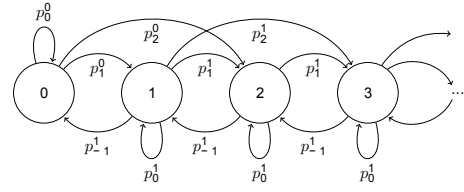}
	\caption[]{The DTMC model for the two UEs case.}
	 \label{fig:DMTC}
\end{figure}

The evolution of the queue at the relay can be modelled as a discrete time Markov Chain (DTMC), as reported in Fig.~\ref{fig:DMTC}. The terms $p_{k}^{0}$ and $p_{k}^{1}$ are the probabilities that the queue size increases by $k$ packets, in a timeslot, when the queue is empty or not, respectively, and their expressions are reported in Appendix C. Finally, by omitting the details for sake of space, we compute $P(Q=0)$ by considering the Z-transformation of the steady-state distribution vector~\cite{book}:
\begin{equation}\label{eq:Q2}
\begin{aligned} 
P(Q=0)=\frac{p_{-1}^{1}-p_{1}^{1}-2p_{2}^{1}}{p_{-1}^{1}-p_{1}^{1}-2p_{2}^{1}+\lambda_{0}}.
\end{aligned} 
\end{equation}

\section{Throughput Analysis}
\label{sec:Thr}
In this section, we derive the network aggregate throughput, $T$,  for $N$ UEs by generalizing the results obtained in Section \ref{sec:PA}. 
In particular, we distinguish between two cases. First, when the queue is stable, $T$ is given by:
\begin{equation}\label{eq:TT2}
\begin{aligned} 
T=NT_{u}=N(T_{ud}+T_{ur}),
\end{aligned} 
\end{equation}
where $T_{u}$ represents the per-user throughput. This is composed by two terms, $T_{ud}$ and $T_{ur}$, which represent the contributions to $T_{u}$ given by the packets received by the mmAP or by $R$, respectively. 
Second, when the queue at $R$ is unstable, the aggregate throughout is:
\begin{equation}\label{eq:TT}
\begin{aligned} 
T=NT_{ud}+\mu_{r}.
\end{aligned} 
\end{equation}
In particular, the expressions for $T_{ud}$ and $T_{ur}$ can be derived as follows. We indicate with $m$ the number of UEs that interfere and with $i$ the number of those that use $\mathrm{FD}$ transmissions ($m-i$ UEs use the $\mathrm{BR}$ scheme). A certain number, $j$, of $\mathrm{FD}$ interferers transmit to $R$ and $i-j$ to the mmAP.
Therefore, $T_{ud}$ and $T_{ur}$ are given by:
\begin{equation}\label{eq:TUidNf}
\begin{aligned} 
T_{ud}&=\Bigl(1-q_{r}P(Q\neq0)\Bigr)T_{ud}^{0}+q_{r}P(Q\neq0)T_{ud}^{1},
\end{aligned} 
\end{equation}
\begin{align*} 
T_{ur}&=q_{u}q_{uf}q_{ur}\sum_{m=0}^{N-1}\binom{N-1}{m}q_{u}^{m}\overline{q}_{u}^{N-1-m}\\
&\times \sum_{i=0}^{m}\binom{m}{i}q_{uf}^{i}q_{ub}^{m-i} \sum_{j=0}^{i} \binom{i}{j}q_{ur}^{j}q_{ud}^{i-j}\\
&\times P_{ur/\{j\}^{f},\{m-i\}^{b}}^{f}+\Bigl(1-q_{r}P(Q\neq0)\Bigr)T_{ur}^{0}\\
&+q_{r}P(Q\neq0)T_{ur}^{1},\stepcounter{equation}\tag{\theequation}\label{eq:TUirNf}
\end{align*} 
where $P(Q=0)$, derived by following the same method used in Section~\ref{sec:PA}, but for $N$ UEs, is given by:
\begin{equation}\label{eq:QN}
\begin{aligned} 
P(Q=0)=\frac{p_{-1}^{1}-\sum_{k=1}^{N}kp_{k}^{1}}{p_{-1}^{1}-\sum_{k=1}^{N}kp_{k}^{1}+\lambda_{0}}.
\end{aligned} 
\end{equation}
In this case, $p_{k}^{1}$, $p_{-1}^{1}$ and $\lambda_{0}$ have the same meaning as for the two UEs case, but different values.
The terms $T_{ud}^{0}$ and $T_{ud}^{1}$ represent the contribution to $T_{u}$ given by the packets sent to the mmAP (when $R$ is interfering or not) and are given by:
\begin{align*} 
T_{ud}^{0}&=q_{u}q_{uf}q_{ud}\sum_{m=0}^{N-1}\binom{N-1}{m}q_{u}^{m}\overline{q}_{u}^{N-1-m} \sum_{i=0}^{m}\binom{m}{i}q_{uf}^{i}q_{ub}^{m-i}\\
&\times\sum_{j=0}^{i} \binom{i}{j}q_{ur}^{j}q_{ud}^{i-j}P_{ud/\{i-j\}^{f},\{m-i\}^{b}}^{f}\\
&+q_{u}q_{ub}\sum_{m=0}^{N-1}\binom{N-1}{m}q_{u}^{m}\overline{q}_{u}^{N-1-m} \sum_{i=0}^{m}\binom{m}{i}q_{uf}^{i}q_{ub}^{m-i}\\
&\times\sum_{j=0}^{i} \binom{i}{j}q_{ur}^{j}q_{ud}^{i-j}\times P_{ud/\{i-j\}^{f},\{m-i\}^{b}}^{b},\stepcounter{equation}\tag{\theequation}\label{eq:Tid0}
\end{align*} 
\begin{align*} 
T_{ud}^{1}&=q_{u}q_{uf}q_{ud}\sum_{m=0}^{N-1}\binom{N-1}{m}q_{u}^{m}\overline{q}_{u}^{N-1-m}\sum_{i=0}^{m}\binom{m}{i}q_{uf}^{i}q_{ub}^{m-i}\\
&\times\sum_{j=0}^{i} \binom{i}{j}q_{ur}^{j}q_{ud}^{i-j}P_{ud/\{i-j,r\}^{f},\{m-i\}^{b}}^{f}\\
&+q_{u}q_{ub}\sum_{m=0}^{N-1}\binom{N-1}{m}q_{u}^{m}\overline{q}_{u}^{N-1-m}\sum_{i=0}^{m}\binom{m}{i}q_{uf}^{i}q_{ub}^{m-i}\\
&\times\sum_{j=0}^{i} \binom{i}{j}q_{ur}^{j}q_{ud}^{i-j}P_{ud/\{i-j,r\}^{f},\{m-i\}^{b}}^{b}.\stepcounter{equation}\tag{\theequation}\label{eq:Tid1}
\end{align*}

Finally, we derive the terms $T_{ur}^{0}$ and $T_{ur}^{1}$:
\begin{align*}
T_{ur}^{0}&=q_{u}q_{ub}\sum_{m=0}^{N-1}\binom{N-1}{m}q_{u}^{m}\overline{q}_{u}^{N-1-m}\\
&\times \sum_{i=0}^{m}\binom{m}{i}q_{uf}^{i}q_{ub}^{m-i} \sum_{j=0}^{i} \binom{i}{j}q_{ur}^{j}q_{ud}^{i-j}\\
&\times P_{ur/\{j\}^{f},\{m-i\}^{b}}^{b}\overline{P}_{ud/\{i-j\}^{f},\{m-i\}^{b}}^{b}.\stepcounter{equation}\tag{\theequation}\label{eq:Tir0}
\end{align*} 
\begin{align*}
T_{ur}^{1}&=q_{u}q_{ub}\sum_{m=0}^{N-1}\binom{N-1}{m}q_{u}^{m}\overline{q}_{u}^{N-1-m}\\
&\times \sum_{i=0}^{m}\binom{m}{i}q_{uf}^{i}q_{ub}^{m-i} \sum_{i=0}^{i} \binom{i}{j}q_{ur}^{j}q_{ud}^{i-j}\\
&\times P_{ur/\{j\}^{f},\{m-i\}^{b}}^{b}\overline{P}_{ud/\{i-j,r\}^{f},\{m-i\}^{b}}^{b}.\stepcounter{equation}\tag{\theequation}\label{eq:Tir1}
\end{align*} 

\section{Numerical Results}
\label{sec:Res}
In this section, we provide the numerical evaluation of the analysis derived in the previous sections. In order to compute the LOS and NLOS probabilities and the path loss, we use the 3GPP model for urban micro cells in outdoor street canyon environment~\cite{3GPP}. More precisely, the path loss depends on the height of the mmAP, $10$ m, the height of the UE, $1.5$ m, the carrier frequency, $f_{c} = 30$ GHz and the distance between the transmitter and the receiver. The transmit and the noise power are set to $P_{t} = 24$~dBm and $P_{N} = -80$~dBm, respectively. Then, the SINR in~\eqref{eq:SINR} and the success probability in~\eqref{eq:prob_succA} are numerically computed. 

\begin{figure}[tb]
	\centering
	\includegraphics[width=7.8cm]{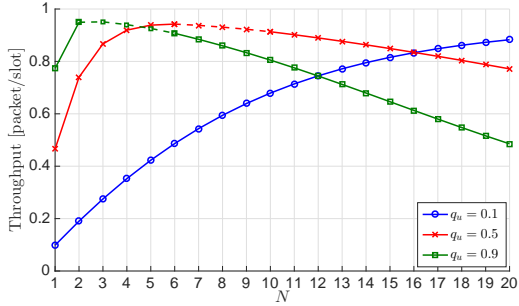}
	\caption[]{$T$ while varying $N$ for several values of $q_{u}$, with $\theta_{rd}=30^\circ $ and $q_{ur} = 0.5$. Solid and dotted lines are used for the range of UEs where the queue is stable or unstable, respectively.}
	 \label{fig:ThrNPerU}
\end{figure}

Hereafter, we show the network throughput ($T$) while varying several parameters. Unless otherwise specified, we set $d_{ur} = 30$ m, $d_{ud} = 50$ m, $\gamma = 10$ dB and $\alpha = 0.1$. Moreover, we set either $\theta_{BW}=5^\circ$ or $\theta_{BW}=\theta_{rd}$ for the $\mathrm{FD}$ and $\mathrm{BR}$ transmissions, respectively. 
In Fig.~\ref{fig:ThrNPerU}, we show $T$ while varying the number of UEs for several UE transmit probability values, i.e., $q_{u}$. In particular, we use solid lines when the queue at $R$ is stable (cf. Eq.~\eqref{eq:TT2}) and the dotted lines when the queue is unstable (cf. Eq.~\eqref{eq:TT}). For $q_{u} = 0.1$ the queue is always stable, in contrast, for $q_{u}= 0.5$ and $q_{u}= 0.9$ the queue becomes unstable at $N= 7$ and $N= 3$, respectively. 
Above a certain number of UEs, $T$ reaches almost the maximum value and then it start decreasing. Namely, for $q_{u}= 0.5$ and $q_{u}= 0.9$, the queue becomes again stable at $N=10$ and $N=6$, respectively, because high values of $N$ and $q_{u}$ lead to high interference that decreases the number of packets successfully received by $R$ and the mmAP. 

\begin{figure}[tb]
	\centering
	\includegraphics[width=7.8cm]{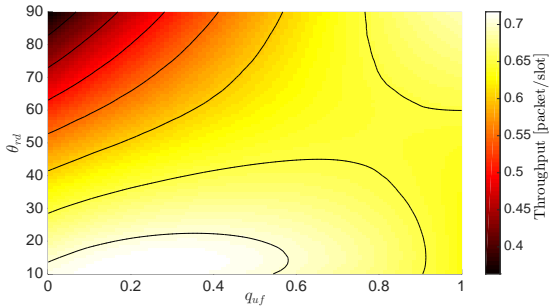}
	\caption[]{$T$ while varying $q_{uf}$ and $\theta_{rd}$ for $N=10$ and $q_{ur}= 0.5$.}
	 \label{fig:ThrQufBW}
\end{figure}

In Fig.~\ref{fig:ThrQufBW}, we show the $T$ while varying the probability of using the $\mathrm{FD}$ scheme, $q_{uf}$, and $\theta_{rd}$. Hereafter, we set $q_{u}= 0.1$ and $N=10$ and we can observe that the optimal choice of $q_{uf}$ depends on $\theta_{rd}$. Namely, for small values of $\theta_{rd}$, $\mathrm{BR}$ transmissions are more preferable, which correspond to small values of $q_{uf}$. Indeed, in this case, we can use beams with high beamforming gain to transmit simultaneously to $R$ and the mmAP. In contrast, for higher values of $\theta_{rd}$, the optimal value of $q_{uf}$ is~$1$ that corresponds to always use the $\mathrm{FD}$ scheme. Furthermore, it is possible to observe that for  $q_{uf}=1$,  $T$ increases with $\theta_{rd}$. This is caused by the interference of $R$ on the communications between the UEs and the mmAP.

\begin{figure}[tb]
	\centering
	\includegraphics[width=7.8cm]{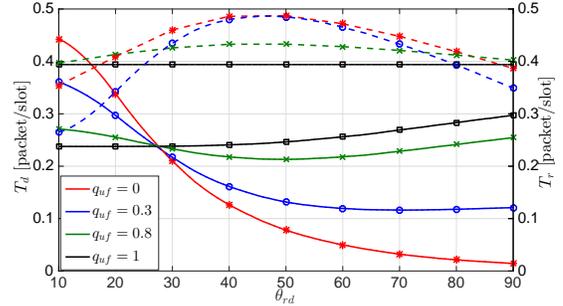}
	\caption[]{$T_{d}$ (solid lines) and $T_{r}$ (dotted lines) with varying $\theta_{rd}$ for several values of $q_{uf}$ and $q_{ur}= 0.5$.}
	 \label{fig:TDTRBW}
\end{figure}

\begin{figure}[tb]
	\centering
	\includegraphics[width=7.8cm]{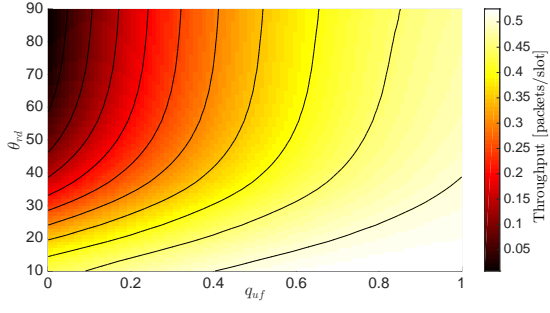}
	\caption[]{$T$ while varying $q_{uf}$ and $\theta_{rd}$ for $N=10$, $q_{ur}= 0.5$ and $\gamma=20$ dB.}
	 \label{fig:ThrQurBW}
\end{figure}

This phenomenon can be better observed in Fig.~\ref{fig:TDTRBW}, which shows both the aggregate throughput received by the mmAP and by $R$, i.e., $T_{d}$ and $T_{r}$, for several values of $q_{uf}$ while varying $\theta_{rd}$. Larger values of $\theta_{rd}$ correspond to longer distances between $R$ and the mmAP, i.e., $d_{rd}$. For $q_{uf}=1$, the success probability for a packet transmitted from $R$ to the mmAP, and so $T_{r}$ (dotted lines), are barely affected by increasing the link length. Indeed, the link R-mmAP is always in LOS. In contrast $T_{d}$ (solid lines) increases for wider $\theta_{rd}$ because the interference caused by $R$ decreases. For $0<q_{uf}<1$, $T_{d}$ and $T_{r}$ have a non-monotonic behavior. Initially, as $\theta_{rd}$ increases, $T_{d}$ decreases because of two reasons. First, the beamforming gain of the $\mathrm{BR}$ transmissions decreases, and so the success probability for a packet sent by using the $\mathrm{BR}$ scheme. Second, since the packets that are not successfully received by the mmAP may increase the number of packets in the queue at $R$, both $T_{r}$ and the interference at the receiver side of the mmAP (caused by the relay) also increase. 
However, above a certain value of $\theta_{rd}$, $T_{r}$ starts decreasing because wider beams with lower beamforming gains are not enough to overcome the path loss. Fig.~\ref{fig:ThrQurBW} shows similar results of Fig.~\ref{fig:ThrQufBW}, but with a higher SINR threshold, i.e., $\gamma=20$ dB. In this case, we can observe that the best transmission strategy is always the $\mathrm{FD}$ scheme, even for low value of $\theta_{rd}$. The reason behind is that the beamforming gain provided by the $\mathrm{BR}$ scheme leads to low success probabilities with respect to the $\mathrm{FD}$ transmission.

\begin{figure}[tb]
	\centering
	\includegraphics[width=7.8cm]{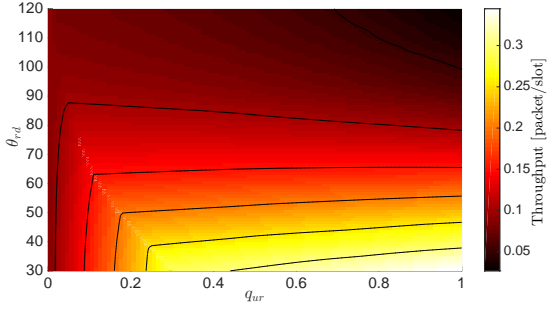}
	\caption[]{$T$ while varying $\theta_{rd}$ and  $q_{ur}$ for $q_{uf}= 1$, $d_{ur}$= 50 m and $d_{ud}= 200$ m.}
	 \label{fig:ThrQurBW2}
\end{figure}

To give further insights into the $\mathrm{FD}$  scheme, we fix $q_{uf}=1$, i.e., UEs always use the FD scheme, and increase the distances, i.e., $d_{ur}=50$ m and $d_{ud}=200$ m. In Fig.~\ref{fig:ThrQurBW2}, we show $T$  when vary $\theta_{rd}$ and $q_{ur}$, which is the probability to transmit to the relay.  In contrast to the previous case (Fig.~\ref{fig:ThrQufBW} and Fig.~\ref{fig:ThrQurBW}), $T$ decreases as $\theta_{rd}$ increases. Indeed, as $d_{rd}$ increases, the high link path loss between $R$ and the mmAP reduces the success probability for a packet sent from $R$ to the mmAP. This has mainly two effects: i) it decreases the interference of $R$ on the communications between the UEs and the mmAP and ii) it reduces the relay's service rate $\mu_{r}$, which makes the queue at $R$ not stable when $q_{ur}$ is above certain values (which is $q_{ur}=0.3$ for $\theta_{rd}=30^\circ$ and decreases as $\theta_{rd}$ increases). Furthermore, we can also observe that for low values of $d_{rd}$, hence $\theta_{rd}$, the highest throughput is given by $q_{ur}=1$, whereas increasing the value of $d_{rd}$, hence $\theta_{rd}$, it is better to always send packets to the relay, i.e., $q_{ur}=0$.

\begin{figure}[tb]
	\centering
	\includegraphics[width=7.8cm]{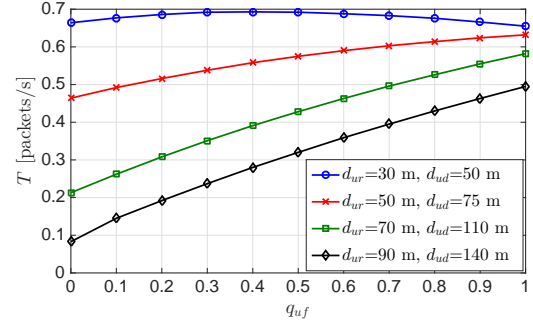}
	\caption[]{$T$ while varying $q_{uf}$ for several values of $d_{ur}$ and $d_{ud}$, when $\theta_{rd}=30$ and $q_{ur}= 0.5$.}
	 \label{fig:ThrQufD}
\end{figure}

Finally, in Fig.~\ref{fig:ThrQufD} we show $T$ while varying $q_{uf}$ for several values of $d_{ur}$ and $d_{ud}$, when $\theta_{rd}=30^\circ$. It is possible to observe that for short distances (blue curve), the optimal value of $q_{uf}$ is smaller than $0.5$. Indeed due to the small path loss values of the links UE-mmAP and UE-$R$, it is always favorable to use the $\mathrm{BR}$ scheme. In contrast, when the distances increase, the transmissions need higher beamforming gain and therefore the $\mathrm{FD}$ scheme is preferable.

\section{Conclusion}
\label{sec:Conc}
In this work, we have presented a throughput analysis for relay assisted mm-wave wireless networks,  where the UEs can transmit by using either a $\mathrm{FD}$ or a $\mathrm{BR}$ transmission. In particular, we have evaluated the performance of the queue at the relay by deriving the stability conditions as well as the arrival and service rates. The numerical evaluation shows that the interference caused by the relay and the link path loss represent the main impediments for the success probability, hence the throughput, in case of short and long distances among the nodes, respectively. Furthermore, results show that the optimal transmission strategy (values of $q_{uf}$ and $q_{ur}$) highly depends on the network topology, e.g., $d_{ud}$, $d_{ur}$ and~$\theta_{rd}$. 

As expected, it is not always beneficial to use narrow beams ($\mathrm{FD}$) compared to wider beams ($\mathrm{BR}$). As a matter of fact, for short distances  and beamwidth of $30^{\circ}$, a $\mathrm{BR}$ transmission is preferable, although it provides a lower beamforming gain. When the distances or the SINR threshold increase, then the $\mathrm{FD}$ scheme should be chosen. Future work will investigate the behavior of the throughput as well as of the delay when taking additional aspects into account, such as the inter-beam interference cancellation technique and the  beamforming alignment phase.


\appendices
\section{}
\label{sec:SPE}
Here, we report the expression for the success probability for the link $ij$ with $N$ symmetric UEs, conditioned to the sets $\mathcal{I}_{f}$ and $\mathcal{I}_{b}$. We average over all the possible scenarios for the LOS and NLOS links. We consider that $k$ and $h$ UEs over $|\mathcal{I}_{f}|$ and $|\mathcal{I}_{b}|$ interferers, respectively, are in LOS. Thus, the success probability is as follows:
\begin{align*} 
&P_{ij/\mathcal{I}_{f},\mathcal{I}_{b}}^{f}=P(\text{LOS}_{ij})P(\text{SINR}_{ij/\mathcal{I}_{f},\mathcal{I}_{b}}^{f} \ge \gamma |\text{LOS}_{ij})\\
&+P(\text{NLOS}_{ij})P(\text{SINR}_{ij/\mathcal{I}_{f},\mathcal{I}_{b}}^{f} \ge \gamma | \text{NLOS}_{ij})\\
&=P(\text{LOS}_{ij})\Biggr[\sum_{k=0}^{|\mathcal{I}_{f}|}\binom{|\mathcal{I}_{f}|}{k}P(\text{LOS}_{ij})^{k}P(\text{NLOS}_{ij})^{|\mathcal{I}_{f}|-k}\\
&\times \sum_{h=0}^{|\mathcal{I}_{b}|}\binom{|\mathcal{I}_{b}|}{h}P(\text{LOS}_{ij})^{h}P(\text{NLOS}_{ij})^{|\mathcal{I}_{b}|-h}\stepcounter{equation}\tag{\theequation}\label{eq:prob_succA}\\
&\times P(\text{SINR}_{ij/\mathcal{I}_{fl},\mathcal{I}_{fn},\mathcal{I}_{bl},\mathcal{I}_{bn}}^{f} \ge \gamma |\text{LOS}_{ij})\Biggr]\\
&+P(\text{NLOS}_{ij})\Biggr[\sum_{k=0}^{|\mathcal{I}_{f}|}\binom{|\mathcal{I}_{f}|}{k}P(\text{LOS}_{ij})^{k}P(\text{NLOS}_{ij})^{|\mathcal{I}_{f}|-k}\\
&\times \sum_{h=0}^{|\mathcal{I}_{b}|}\binom{|\mathcal{I}_{b}|}{h}P(\text{LOS}_{ij})^{h}P(\text{NLOS}_{ij})^{|\mathcal{I}_{b}|-h}\\
&\times P(\text{SINR}_{ij/\mathcal{I}_{fl},\mathcal{I}_{fn},\mathcal{I}_{bl},\mathcal{I}_{bn}}^{f} \ge \gamma |\text{NLOS}_{ij})\Biggr].
\end{align*} 

The expressions $P(\text{SINR}_{ij/\mathcal{I}_{f},\mathcal{I}_{b}}^{f} \ge \gamma |\text{LOS}_{ij})$ and $P(\text{SINR}_{ij/\mathcal{I}_{f},\mathcal{I}_{b}}^{f} \ge \gamma | \text{NLOS}_{ij})$ are the probabilities, conditioned to the specific scenarios of interferers, $\mathcal{I}_{f}$ and $\mathcal{I}_{b}$,  that the received SINR is above $\gamma$, when link $ij$ is in LOS and NLOS, respectively. 

\section{}
\label{sec:QA}
In this appendix, we report the terms $A_{r}$ and $B_{r}$, which are used in Section~\ref{sec:PA} for the expressions of $\lambda_{1}$ and $\mu_{r}$, respectively, and can be computed similarly to $\lambda_{0}$:
\begin{align*}
A_{r}&=2q_{u}\overline{q}_{u}q_{uf}q_{ur}P_{ur}^{f}+2q_{u}\overline{q}_{u}q_{ub}P_{ur}^{b}\overline{P}_{ud}^{b}\\
&+q_{u}^2q_{uf}^2q_{ur}^2q_{ur}^2\Bigl[2P_{ur/\{1\}^{f}}^{f}\overline{P}_{ur/\{1\}^{f}}^{f}\\
&+2\Bigl(P_{ur/\{1\}^{f}}^{f}\Bigl)^{2}\Bigl]+2q_{u}^{2}q_{uf}^{2}q_{ur}q_{ud}P_{ur}^{f}\stepcounter{equation}\tag{\theequation}\label{eq:AR}\\
&+2q_{u}^{2}q_{1f}q_{ub}q_{ur}\Bigl[P_{ur/\{1\}^{b}}^{f}\Bigl(1-P_{ur/\{1\}^{f}}^{b}\overline{P}_{ud/\{r\}^{f}}^{b}\Bigl)\\
&+\overline{P}_{ur/\{1\}^{b}}^{f}P_{ur/\{1\}^{f}}^{b}\overline{P}_{ud/\{r\}^{f}}^{b}+2\Bigl(P_{ur/\{1\}^{f}}^{b}\overline{P}_{ud/\{r\}^{f}}^{b}\Bigl)^{2}\Bigl]\\
&+2q_{u}^{2}q_{ub}q_{uf}q_{ud}P_{ur}^{b}\overline{P}_{ud/\{1,r\}^{f}}^{b}+q_{u}^{2}q_{ub}^{2}\\
&\times\Bigl[2P_{ur/\{1\}^{b}}^{b}\overline{P}_{ud/\{r\}^{f},\{1\}^{b}}^{b}\Bigl(1-P_{ur/\{1\}^{b}}^{b}\overline{P}_{ud/\{r\}^{f},\{1\}^{b}}^{b}\Bigl)\\
&+2\Bigl(P_{ur/\{1\}^{b}}^{b}\overline{P}_{ud/\{r\}^{f},\{1\}^{b}}^{b}\Bigl)^{2}\Bigl].
\end{align*}
\begin{align*}
B_{r}&=P_{rd}^{f}\Bigl(\overline{q}_{u}^{2}+2q_{u}\overline{q}_{u}q_{uf}q_{ur}+q_{u}^{2}q_{uf}^{2}q_{2f}q_{ur}^{2}\Bigl)\\
&+P_{rd/\{1\}^{f}}^{f}\Bigl(2q_{u}\overline{q}_{u}q_{uf}q_{ud}+2q_{u}^{2}q_{uf}^{2}q_{ud}q_{ur}\Bigl)\\
&+P_{rd/\{1\}^{b}}^{f}\Bigl(2q_{u}\overline{q}_{u}q_{ub}+2q_{u}^{2}q_{ub}q_{uf}q_{ur}\Bigl)\stepcounter{equation}\tag{\theequation}\label{eq:muR2}\\
&+P_{rd/\{2\}^{f}}^{f}q_{u}^{2}q_{uf}^{2}q_{ud}^{2}+P_{rd/\{1\}^{f},\{1\}^{b}}^{f}2q_{u}q_{uf}q_{ub}q_{ud}\\
&+P_{rd/\{2\}^{b}}^{f}q_{u}^{2}q_{ub}^{2}.
\end{align*}

\section{}
\label{sec:TP}
Hereafter, we present the transition probabilities $p_{k}^{0}$ and $p_{k}^{1}$ for the two UEs case. 
\begin{align*} 
p_{-1}^{1}&=q_{r}\Bigl[P_{rd}^{f}\Bigl(\overline{q}_{u}^{2}+2q_{u}\overline{q}_{u}q_{uf}q_{ur}\overline{P}_{ur}^{f}\\
&+(q_{u}q_{uf}q_{ur}\overline{P}_{ur/\{1\}^{f}}^{f})^{2}\Bigl)+P_{rd/\{1\}^{f}}^{f}\\
&\times \Bigl(2q_{u}\overline{q}_{u}q_{uf}q_{ud}+2q_{1}^{2}q_{uf}^{2}q_{ud}q_{ur}\overline{P}_{ur}^{f}\Bigl)\stepcounter{equation}\tag{\theequation}\label{eq:p_1f}\\
&+P_{rd/\{1\}^{b}}^{f}\Bigl(2q_{u}\overline{q}_{u}q_{ub}(1-P_{ur}^{b}\overline{P}_{ud/\{r\}^{f}}^{b})\\
&+2q_{u}^{2}q_{ub}q_{uf}q_{ur}(1-P_{ur/\{1\}^{f}}^{b}\overline{P}_{ud/\{r\}^{f}}^{b})\overline{P}_{ur/\{1\}^{b}}^{f}\Bigl)\\
&+P_{rd/\{1\}^{f},\{1\}^{b}}^{f}2q_{u}^{2}q_{uf}q_{ub}q_{ud}(1-P_{ur}^{b}\overline{P}_{ud/\{1,r\}^{f}}^{b})\\
&+P_{rd/\{2\}^{b}}^{f}\Bigl(q_{u}q_{ub}(1-P_{ur/\{1\}^{b}}^{b}\overline{P}_{ud/\{r\}^{f},\{1\}^{b}}^{b})\Bigl)^{2}\Bigl]\\
&+P_{rd/\{2\}^{f}}^{f}q_{u}^{2}q_{uf}^{2}q_{ud}^{2}.
\end{align*} 
\begin{align*} 
p_{1}^{0}&=2q_{u}\overline{q}_{u}q_{uf}q_{ur}P_{ur}^{f}+2q_{u}\overline{q}_{u}q_{ub}P_{ur}^{b}\overline{P}_{ud}^{b}\\
&+2q_{u}^{2}q_{uf}^{2}q_{ur}^{2}P_{ur/\{1\}^{f}}^{f}\overline{P}_{ur/\{1\}^{f}}^{f}+2q_{u}^{2}q_{uf}^{2}q_{ur}q_{ud}P_{ur}^{f}\\
&+2q_{u}^{2}q_{uf}q_{ub}q_{ur}\Bigl[P_{ur/\{1\}^{b}}^{f}\Bigl(1-P_{ur/\{1\}^{f}}^{b}\overline{P}_{ud}^{b}\Bigl)\stepcounter{equation}\tag{\theequation}\label{eq:p01b}\\
&+\overline{P}_{ur/\{1\}^{b}}^{f}P_{ur/\{1\}^{f}}^{b}\overline{P}_{ud}^{b}\Bigl]+2q_{u}^{2}q_{ub}q_{uf}q_{ud}P_{ur}^{b}\overline{P}_{ud/\{1\}^{f}}^{b}\\
&+q_{1}^{2}q_{ub}^{2}\Bigl[2P_{ur/\{1\}^{b}}^{b}\overline{P}_{ud/\{1\}^{b}}^{b}\Bigl(1-P_{ur/\{1\}^{b}}^{b}\overline{P}_{ud/\{1\}^{b}}^{b}\Bigl)\Bigl].
\end{align*} 
\begin{align*} 
p_{1}^{1}&=\overline{q}_{r}p_{1}^{0}+q_{r}\Bigl[2q_{u}\overline{q}_{u}q_{uf}q_{ur}P_{ur}^{f}\overline{P}_{rd}^{f}\\
&+2q_{u}\overline{q}_{u}q_{ub}P_{ur}^{b}\overline{P}_{ud/\{r\}^{f}}^{b}\overline{P}_{rd/\{1\}^{b}}^{f}\\
&+2q_{1}^{2}q_{uf}^{2}q_{ud}q_{ur}P_{ur}^{f}\overline{P}_{rd/\{1\}^{f}}^{f}\\
&+2q_{u}^{2}q_{uf}q_{ub}q_{ud}P_{ur}^{b}\overline{P}_{ud/\{1,r\}^{f}}^{b}\overline{P}_{rd/\{1\}^{f},\{1\}^{b}}^{f}\\
&+q_{u}^{2}q_{uf}^{2}q_{ur}^{2}\Bigl(P_{ur/\{1\}^{f}}^{f}\overline{P}_{ur/\{1\}^{f}}^{f}\overline{P}_{rd}^{f}+(P_{ur/\{1\}^{f}}^{f})^{2}P_{rd}^{f}\Bigl)\\
&+q_{u}^{2}q_{ub}^{2}\Bigl(2P_{ur/\{1\}^{b}}^{b}\overline{P}_{ud/\{r\},\{1\}}^{b}\overline{P}_{rd/\{2\}^{b}}^{f}\\
&\times(1-P_{ur/\{1\}^{b}}^{b}\overline{P}_{ud/\{r\}^{f},\{1\}^{b}}^{b})\stepcounter{equation}\tag{\theequation}\label{eq:p11f}\\
&+(P_{ur/\{1\}^{b}}^{b}\overline{P}_{ud/\{r\}^{f},\{1\}^{b}}^{b})^{2} {P}_{rd/\{2\}^{b}}^{f}\Bigl)\\
&+2q_{u}^{2}q_{ub}q_{uf}q_{ur}\Bigl(P_{ur/\{1\}^{f}}^{b}\overline{P}_{ud/\{r\}^{f}}^{b}\overline{P}_{ur/\{r\}^{f},\{1\}^{b}}^{f}\overline{P}_{rd/\{1\}^{b}}^{f}\\
&+(1-P_{ur/\{1\}^{f}}^{b}\overline{P}_{ud/\{r\}^{f}}^{b})P_{ur/\{1\}^{b}}^{f}\overline{P}_{rd/\{1\}^{b}}^{f}\\
&+P_{ur/\{2\}^{f}}^{b}\overline{P}_{ud/\{r\}^{f}}^{b}P_{ur/\{1\}^{b}}^{f}P_{rd/\{1\}^{b}}^{f}\Bigl)\Bigl].
\end{align*} 
\begin{align*} 
p_{2}^{0}&=\Bigl(q_{u}q_{uf}q_{ur}P_{ur/\{1\}^{f}}^{f}\Bigl)^{2}+\Bigl(q_{u}q_{ub}P_{ur/\{1\}^{b}}^{b}\overline{P}_{ud/\{r\}^{f},\{1\}^{b}}^{b}\Bigl)^{2}\\
&+2q_{1}^{2}q_{ub}q_{uf}q_{ur}P_{ur/\{1\}^{f}}^{b}\overline{P}_{ud}^{b}P_{ur/\{1\}^{b}}^{f}\stepcounter{equation}\tag{\theequation}\label{eq:p02}.
\end{align*} 
\begin{align*} 
p_{2}^{1}&=\overline{q}_{r}p_{2}^{0}+q_{r}\Bigl[\Bigl(q_{u}q_{uf}q_{ur}P_{ur/\{1\}^{f}}^{f}\Bigl)^{2}\overline{P}_{rd}^{f}\\
&+\Bigl(q_{u}q_{ub}P_{ur/\{1\}^{b}}^{b}\overline{P}_{ud/\{r\}^{f},\{1\}^{b}}^{b}\Bigl)^{2}\overline{P}_{rd/\{2\}^{b}}^{f}\stepcounter{equation}\tag{\theequation}\label{eq:p12}\\
&+2q_{1}^{2}q_{ub}q_{uf}q_{ur}P_{ur/\{1\}^{f}}^{b}\overline{P}_{ud}^{b}P_{ur/\{1\}^{b}}^{f}\overline{P}_{rd/\{1\}^{b}}^{f}\Bigl].
\end{align*} 

\bibliography{ref}
\bibliographystyle{IEEEtran}

\end{document}